\documentclass[fleqn,10pt]{wlscirep}

\usepackage{bm}
\usepackage{comment}
\usepackage{hyperref}
\usepackage[symbol]{footmisc}

\title{Level spacing statistics for light in two-dimensional disordered photonic
crystals
\footnote[2]{This is a pre-print of an article published in \textit{Scientific Reports}. The final authenticated version is available online at: \href{https://doi.org/10.1038/s41598-018-29996-1}{https://doi.org/10.1038/s41598-018-29996-1.}}}

\author[1]{Jose M. Escalante}
\author[1,*]{Sergey E. Skipetrov}
\affil{Univ. Grenoble Alpes, CNRS, LPMMC, 38000 Grenoble, France}

\affil[*]{sergey.skipetrov@lpmmc.cnrs.fr}

\begin{abstract}
We study the distribution of eigenfrequency spacings (the so-called level spacing statistics) for light in a two-dimensional (2D) disordered photonic crystal composed of circular dielectric (silicon) rods in air. Disorder introduces localized transverse-magnetic (TM) modes into the band gap of the ideal crystal. The level spacing statistics is found to approach the Poisson distribution for these modes. In contrast, for TM modes outside the band gap and for transverse-electric (TE) modes at all frequencies, the level spacing statistics follows the Wigner-Dyson distribution.
\end{abstract}

\begin{document}

\flushbottom
\maketitle
%
%
\thispagestyle{empty}


\section*{Introduction}

The optical response of any dielectric medium can be calculated from the eigenmodes $\psi_{n}$ and eigenfrequencies $\omega_n$ of Maxwell equations in the medium, with appropriate boundary conditions. In a disordered medium where the dielectric function $\varepsilon$ varies randomly in space, $\omega_n$ are random numbers that should be analyzed in statistical terms. Whereas the probability distribution of $\omega_n$ can strongly depend on the particular medium under study, the distribution $P(s)$ of normalized nearest-neighbor eigenfrequency \textit{spacings} $s_n = \Delta \omega_n/\langle \Delta \omega_n \rangle$ often takes one of several universal forms independent of details of the medium. Here $\Delta \omega_n = \omega_n  - \omega_{n-1}$, the eigenfrequencies are assumed to be ordered in the ascending order ($\omega_n \geq \omega_{n-1}$), and the angular brackets $\langle \cdots \rangle$ denote ensemble averaging. $P(s)$ first came into light for energy \textit{level} spacings of complex nuclei \cite{wigner51,bohigas83} and is therefore called the ``(nearest-neighbor) level-spacing distribution'' in the physics literature whenever one speaks of either quantum-mechanical systems or eigenfrequencies of Maxwell or other wave equations \cite{mehta67,izrailev90,guhr98,haake01}; we adopt this terminology here as well. For light or any other wave in a disordered medium, the shape of $P(s)$ depends on the spatial extent of the modes $\psi_{n}$. For weak disorder, the modes $\psi_{n}$ are extended in space over the whole disordered sample and the system is classically chaotic. The normalized eigenvalue spacings are then expected to obey the Wigner-Dyson (WD) distribution \cite{mehta67,izrailev90,guhr98,haake01}:
\begin{equation}
P_{\textrm{WD}}(s) = \frac{\pi s}{2} \exp\left(-\frac{\pi s^2}{4} \right).
\label{P_WD}
\end{equation}
This equation turns out to apply to many types of classically chaotic systems, such as, e.g., chaotic billiards \cite{haake01}. Its most important feature is the so-called level-repulsion phenomenon: $P_{\textrm{WD}}(s) \to 0$ for $s \to 0$. The level repulsion is directly related to the extended nature of eigenmodes because the latter should be mutually orthogonal and thus cannot have the same frequency given that they overlap in space. Therefore, it is impossible to have $s=0$. In contrast, the frequencies of two spatially localized modes are uncorrelated if the modes are localized in two distant regions of space and have no spatial overlap. This leads to the Poisson distribution of normalized eigenfrequency spacings \cite{mehta67,izrailev90,guhr98,haake01}:
\begin{equation}
P_{\textrm{P}}(s) = \exp(-s).
\label{P_P}
\end{equation}
The level repulsion is absent here and the frequencies of two modes can be arbitrary close to each other or even coincide: $P_{\textrm{P}}(0) > 0$.

In this paper, we study the level-spacing distribution $P(s)$ for light in a disordered two-dimensional (2D) photonic crystal. Whereas the eigenmodes of the ideal crystal are extended, disorder induces localized eigenmodes due to the phenomenon of Anderson localization \cite{anderson58,lagendijk09}. The latter is widespread in physics and under appropriate circumstances, takes place for electrons in disordered solids \cite{anderson58,kramer93}, atoms in random potentials
\cite{billy08,chabe08,jendr12}, sound \cite{hu08} and light \cite{segev13} in disordered materials. The optical case is of special interest because of numerous existing and emerging technological applications of disordered photonics \cite{wiersma13,redding13,hsieh15,trojak17}. Early theoretical work on Anderson localization of light exploited the analogy between Maxwell and Schr\"{o}dinger equations to adapt to light the results first obtained for electrons in disordered solids \cite{john84,john87,anderson85}. Differences between the two cases were also identified, originating mainly from a different dispersion relation and the resonant nature of scattering for optical waves \cite{john91,vantiggelen94}. Experiments were quite successful in low-dimensional (1D \cite{berry97}, quasi-1D \cite{chabanov00} and 2D \cite{schwartz07}) systems but inconclusive in 3D \cite{wiersma97,vanderbeek12,sperling13,sperling16} where an undisputable evidence of disorder-induced localization is still lacking \cite{skip16}. Recent work suggests that the vector nature of light (i.e., the presence of the polarization degree of freedom) and its associated strong near-field coupling between nearby scatterers by longitudinal fields (the so-called dipole-dipole interactions) play a crucial role for Anderson localization of light in 1D \cite{faez11}, 2D \cite{maximo15}, and 3D \cite{skip14} systems. In particular, they prevent Anderson localization in a simple 3D model in which light is scattered by randomly distributed resonant point scatterers \cite{skip14,bellando14}. Although it is unclear whether the same mechanism counteracts localization of light in other disordered photonic systems (powders of small dielectric or semiconductor particles, porous media, etc.), estimations show that longitudinal near fields in light scattering from a single dielectric sphere are of the same order as for a point-like scatterer \cite{escalante17}.

Disordered photonic crystals have been first proposed as candidate systems for observing Anderson localization by Sajeev John \cite{john87} and later studied by many authors \cite{sapienza10,riboli14,riboli17,crane17,lee18,sigalas96,asatryan99,vanneste05,froufe17,huis12,faggiani16,mazoyer09,vasco17,garcia17}. (These references are only a few among many works dealing with 2D as well as with 3D disordered photonic crystals.) However, the level spacing statistics of these systems has not been studied until now. Localization of light in a disordered 2D photonic crystal presents some particularities that make it quite different from the ``canonical'' case of Anderson localization of a scalar wave in a completely disordered system without an underlying periodic structure. First, the band structures of the ideal crystal are different for light polarized perpendicular to the plane of propagation (transverse-magnetic or TM modes) and in the plane (transverse-electric or TE modes). The impact of disorder on the two types of modes turns out to be different as well. Second, localized states inside the bandgap of the ideal crystal and near the band edges are due to two distinct physical mechanisms: the former are defect modes that require disorder to exist but that are destroyed when the disorder is increased, whereas the latter are genuine Anderson localized modes that shrink with increasing disorder. This is a typical situation in disordered photonic-crystal structures \cite{huis12,faggiani16}. Numerous experimental realizations of 2D disordered photonic crystals exist, allowing for observation of various disorder-related phenomena, including those with a potential for practical applications \cite{sapienza10,riboli14,riboli17,crane17,lee18}.

To study the level-spacing statistics in disordered 2D photonic crystals and its relation with extended or localized nature of electromagnetic modes, we first estimate the localization length $\xi$ of the latter and perform a finite-size scaling analysis by studying the sensitivity of $\xi$ to the size $L$ of the disordered crystal. The independence of $\xi$ from $L$ signals localized modes. For the disordered photonic crystals studied in this work, we can claim the existence of localized modes with certainty only for TM polarization and frequencies inside the band gap of the crystal without disorder. We then analyze the statistics of spacings between eigenfrequencies of Maxwell equations in our 2D systems and establish a correspondence between the level repulsion phenomenon and the extended character of modes. The level spacing distribution is close to the Wigner-Dyson distribution (\ref{P_WD}) in this case. In the opposite case of localized modes, the level spacing distribution approaches the Poison distribution (\ref{P_P}) expected for wave systems with uncorrelated eigenfrequencies.

\section*{\label{sec:model}The model}

In 2D dielectric media, TM- and TE-polarized waves decouple from each other. For a constant magnetic permeability $\mu = 1$ and a random dielectric constant $\epsilon(\bm{\rho})$ with $\bm{\rho} = (x, y)$, the electric field $E_z$ of a TM-polarized monochromatic wave obeys a scalar wave equation
\begin{eqnarray}
\left[ \bm{\nabla}^2 + \frac{\omega^2}{c^2} \epsilon(\bm{\rho}) \right] E_z(\bm{\rho}) = 0,
\label{tmmodes}
\end{eqnarray}
where $\bm{\nabla} = \mathbf{e}_x \partial/\partial x + \mathbf{e}_y \partial/\partial y$. On the other hand, the magnetic field $H_z$ of a monochromatic TE-polarized wave obeys
\begin{eqnarray}
\bm{\nabla} \cdot \left[ \frac{1}{\epsilon(\bm{\rho})} \bm{\nabla} H_z(\bm{\rho}) \right] + \frac{\omega^2}{c^2} H_z(\bm{\rho}) = 0.
\label{temodes}
\end{eqnarray}
The crucial difference between the two cases stems from the fact that in the TE case, the electric field $\mathbf{E}(\bm{\rho})$ is a vector lying the $xy$ plane whereas in the TM case, $E_z(\bm{\rho})$ is a scalar. This trivial observation can have important consequences as far as Anderson localization is concerned. It was found, for example, that only TM modes become spatially localized in a dense 2D random arrangement of identical two-level atoms whereas TE modes remain extended \cite{maximo15}.

\begin{figure*}[t]
\centering
\includegraphics[width=0.8\textwidth]{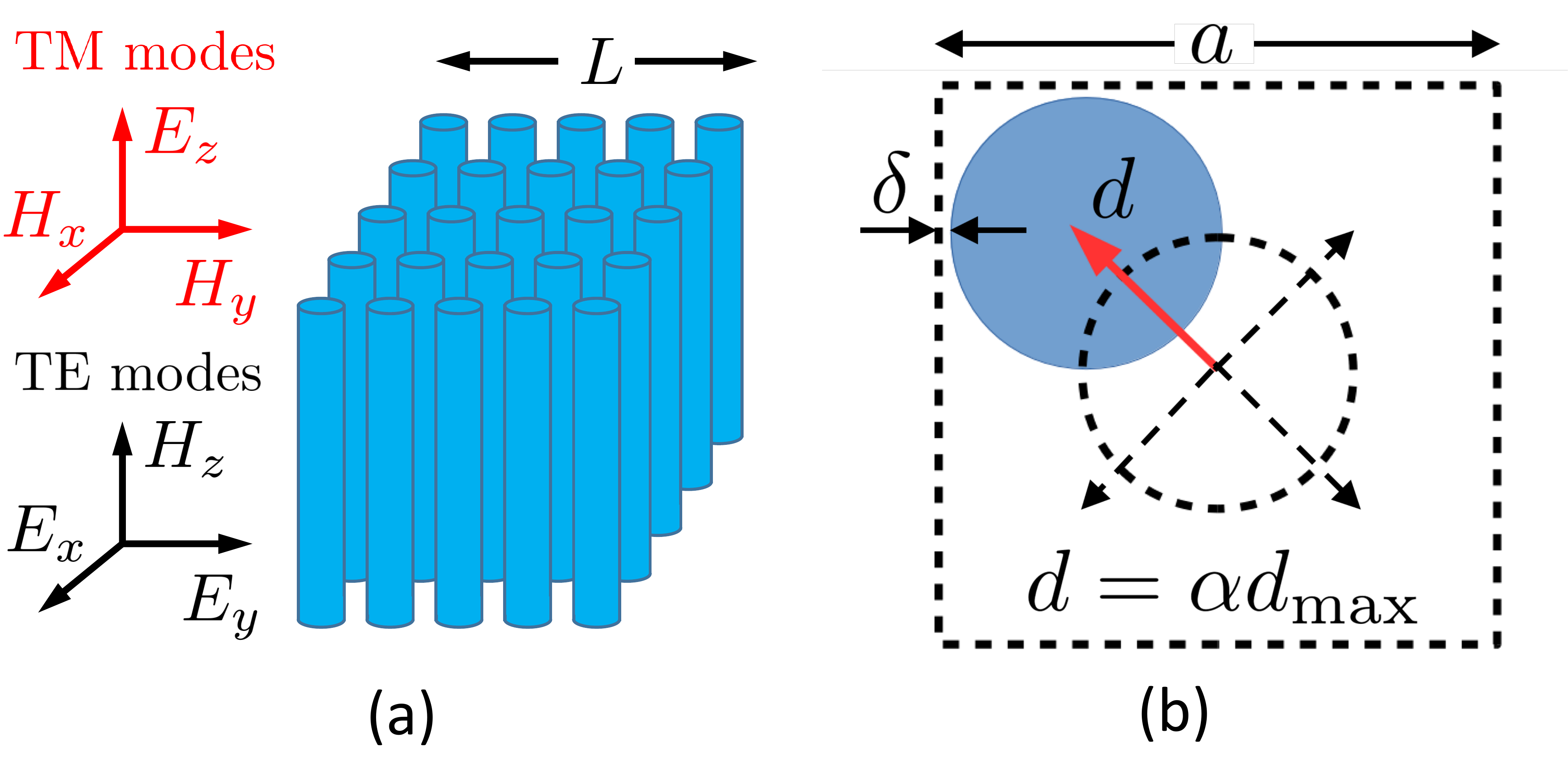}
\caption{(a) The considered physical systems is a 2D square array of dielectric cylinders aligned along the $z$ axis. We impose periodic boundary conditions along the $x$ and $y$ axes. The linear size of the system is denoted by $L$. The TM (TE) modes of the electromagnetic field have the electric (magnetic) field parallel to $z$. (b) Disorder is introduced into the regular array by displacing each cylinder by a distance $d$ in a direction that is randomly chosen among the four diagonal directions inside the unit cell of the periodic structure. The maximum displacement $d_{\mathrm{max}}$ is determined by the requirement that the cylinder does not approach the cell boundary closer than $\delta = \delta_{\mathrm{min}} = 0.02 a$ to avoid computational problems.}
\label{fig_structure}
\end{figure*}

\begin{figure*}[h!]
\centering
\includegraphics[width=0.9\textwidth]{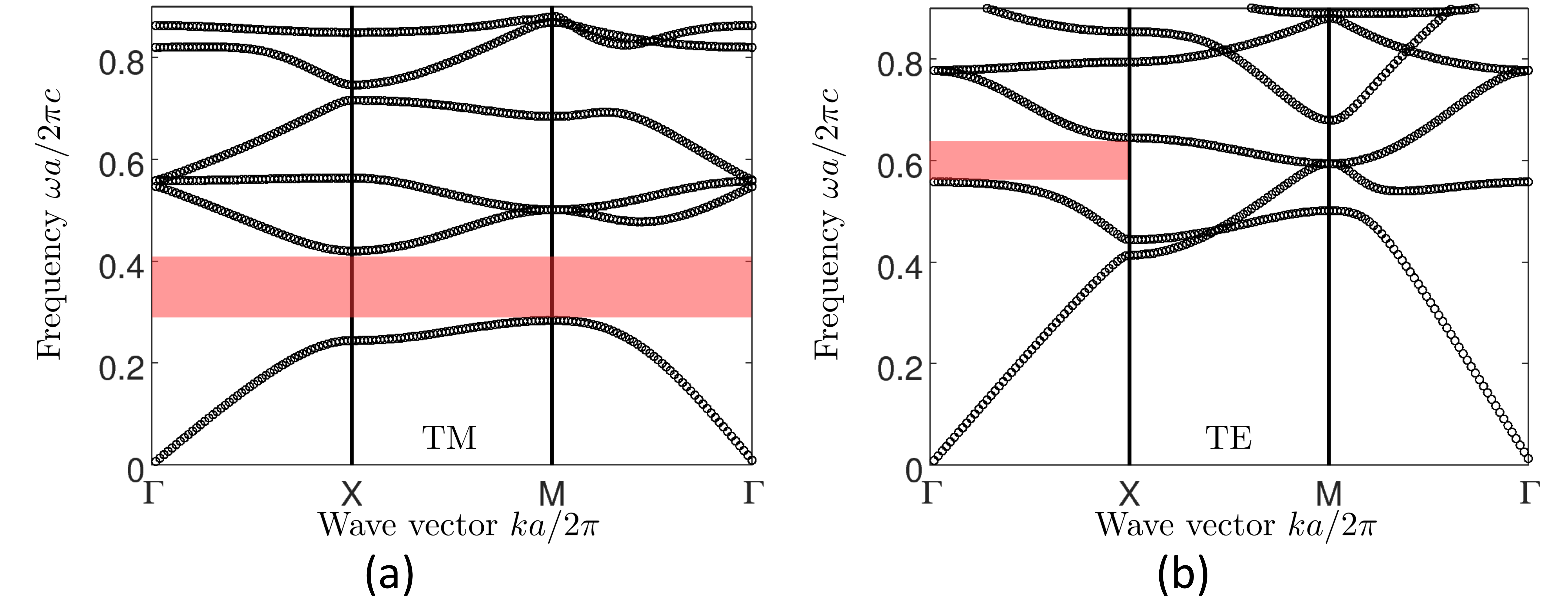}
\caption{TM (a) and TE (b) band structures of the considered photonic crystal without disorder. Pink areas represent the full (for TM polarization) and partial (for TE polarization) band gaps that we focus on in this work.}
\label{fig_band_structure}
\end{figure*}

\begin{figure*}[t]
\centering
\includegraphics[width=0.9\textwidth]{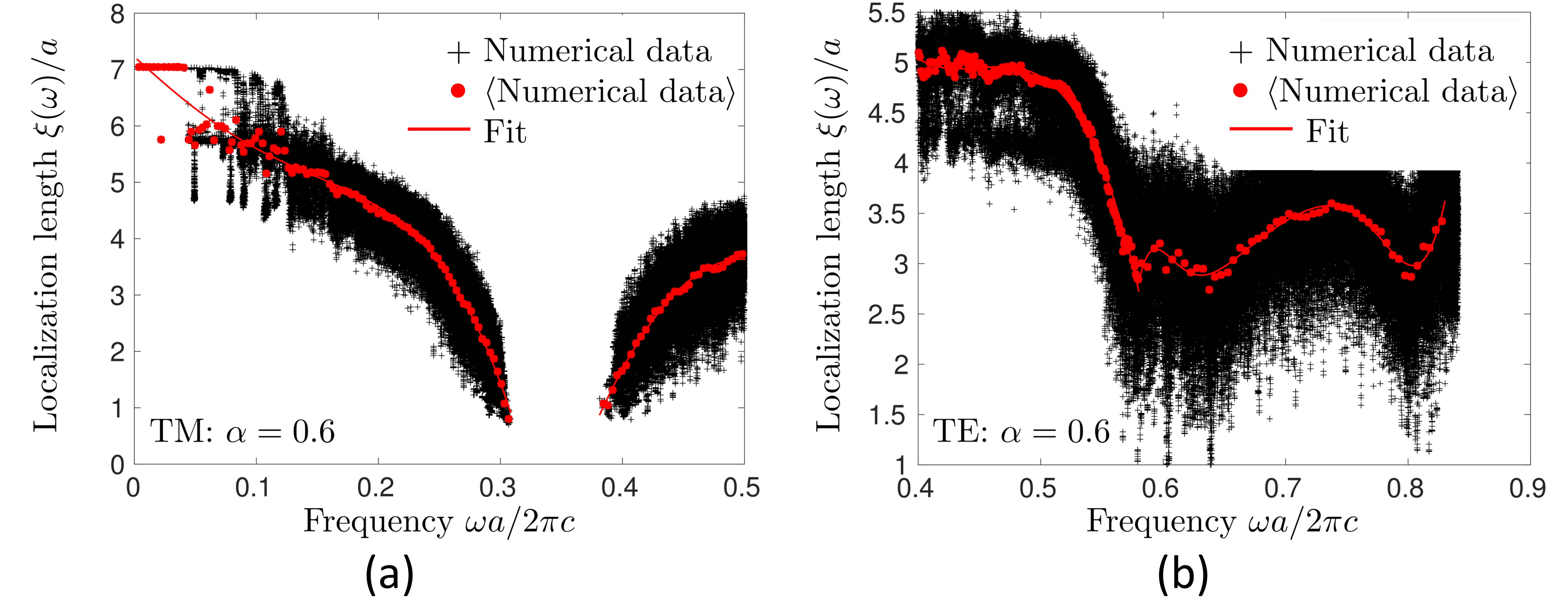}
\caption{Illustration of the method that we employ to determine the average localization length $\langle \xi(\omega) \rangle$ for TM (a) and TE (b) modes (here for $\alpha=0.6$ and $L = 15a$). First, the localization length $\xi_{n,\mathbf{k}}$ is estimated for each mode from equation\ (\ref{LL}), for as much as 50 realizations of disorder (black crosses). Then an average value $\langle \xi(\omega) \rangle$ is obtained by averaging all $\xi_{n,\mathbf{k}}$ corresponding to $\omega_n(\mathbf{k})$ within a narrow frequency band $\Delta \omega = 6 \times 10^{-3}$ around $\omega$ (red dots). Finally, polynomial fits are performed to obtain smooth lines (red solid lines). Note that the fits are performed independently on the left and on the right from the full (for TM polarization) or partial (for TE polarization) band gap, so that the resulting dependence $\langle \xi(\omega) \rangle$ appears to have an unphysical discontinuity.}
\label{fig_method}
\end{figure*}

\begin{figure*}[h!]
\centering
\includegraphics[width=\textwidth]{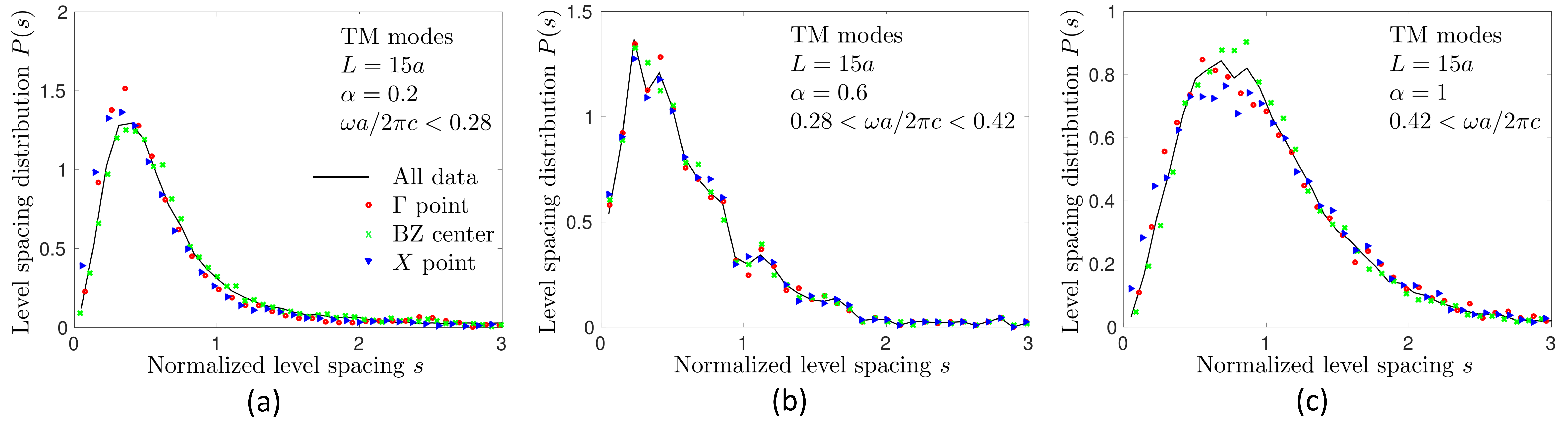}
\caption{Probability density of normalized level spacings $s$ computed by taking into account all wave vectors $\mathbf{k} = \mathbf{e}_x k$ (black solid lines) or only those in the vicinity of the $\Gamma$ point ($\mathbf{k} = 0$, red circles), in the middle of the irreducible Brillouin zone ($\mathbf{k} = \mathbf{e}_x \pi/2L$, green crosses), or in the vicinity of $X$ point ($\mathbf{k} = \mathbf{e}_x \pi/L$, blue triangles). The panel (a) is obtained for $\alpha = 0.2$ and frequencies below the first bandgap of the structure without disorder ($\alpha = 0$), the panel (b) for $\alpha = 0.6$ and frequencies inside the first bandgap of the structure without disorder, and the panel (c) for $\alpha = 1$ and frequencies above the bandgap.}
\label{statistics_k}
\end{figure*}

We consider a square array of infinitely long dielectric cylinders with a dielectric constant $\epsilon = 11.7$ typical for silicon in the infrared \cite{edwards80,deg13}, embedded in air (see figure~\ref{fig_structure}a). The cylinder radius is set to be $r=0.2a$, where $a$ is the lattice parameter. System sizes $L$ up to $15a$ are considered. The infinite structure ($L \to \infty$) has a complete photonic bandgap for TM modes with frequencies $\omega$ in the range $\omega a/2 \pi c \in [0.28, 0.43]$ and a partial bandgap for TE modes in the $\Gamma$X direction for $\omega a/2 \pi c \in [0.55 , 0.66]$ \cite{joan08,deg13}, see figure \ref{fig_band_structure}. These frequency ranges are the most interesting for us because the presence of a spectral gap in the periodic structure indicates strong (destructive) interference effects for light in the dielectric material, creating favorable conditions for the appearance of localized modes upon introducing disorder \cite{john87,john91}. Disorder is introduced in the crystal by displacing each cylinder along a randomly chosen diagonal of the unit cell by a distance $d = \alpha d_{\mathrm{max}}$ (see figure\ \ref{fig_structure}b), where the maximum displacement is $d_{\mathrm{max}} = \sqrt{2}[a/2 - (r+\delta_{\mathrm{min}})]$, ensuring that a cylinder does not approach a cell boundary closer than $\delta = \delta_{\mathrm{min}} = 0.02 a$. The strength of disorder is quantified by a parameter $\alpha \in [0,1]$, with $\alpha = 0$ corresponding to a perfect crystal and $\alpha = 1$ to the most random structure. The resulting structure exhibits complicated structural correlations that are different from those considered in the recent work on Anderson localization and bandgap formation in hyperuniform random structures \cite{froufe17}.

In experiments, 2D photonic crystals are often confined between two parallel reflecting planes perpendicular to the cylinders (or holes) forming the periodic structure \cite{sapienza10,riboli14,riboli17,crane17,lee18,huis12,faggiani16,mazoyer09,vasco17,garcia17}. These systems, often called ``photonic crystal slabs'', support TM- or TE-like modes that can be excited by an appropriate external excitation \cite{joan08}. The modes of the array of infinite cylinders shown in figure \ref{fig_structure}a and those of a photonic crystal slab are equivalent as far as only the in-plane propagation and scattering [wave vector in the $(x,y)$ plane] are concerned. Because some amount of out-of-plane scattering is inevitable in real experimental systems, our results may be considered as an approximation of a realistic structure in which cylinders have finite length and in-plane scattering dominates.

\section*{\label{sec:methods}Computational approach}

Assuming periodic boundary conditions in the $xy$ plane, we consider our disordered system as a unit cell (a ``supercell'') of a 2D unbounded photonic crystal with a lattice constant $L$.  The eigenfunctions of equations\ (\ref{tmmodes}) and (\ref{temodes}) in such a crystal can be written in the form \cite{joan08,lourtioz05}
\begin{eqnarray}
\psi_{n,\mathbf{k}}(\bm{\rho}) = u_{n,\mathbf{k}}(\bm{\rho}) \exp(i \mathbf{k} \bm{\rho}),
\label{period}
\end{eqnarray}
where $\psi$ denotes either $E_z$ or $H_z$ and $u_{n,\mathbf{k}}(\bm{\rho})$ are periodic functions of $x$ and $y$. Sufficiently strong disorder is expected to make the statistical properties of eigenfunctions and eigenfrequencies of the disordered crystal that we are going to study in the following independent from the direction of the wave vector $\mathbf{k}$. It is therefore sufficient to consider $\mathbf{k}$ parallel to the $x$ axis and restricted to a half of the first Brillouin zone $0 \leq k \leq \pi/L$ (the irreducible zone). We use the standard software package FreeFem++ implementing the finite element method \cite{FEM} and employ the Galerkin method \cite{JIA2002} to calculate the band structure $\omega_n(k)$ and the eigenfunctions (modes) $u_{n,\mathbf{k}}(\bm{\rho})$. The localization properties of the modes are characterized by their inverse participation ratio (IPR)
\begin{equation}
\mathrm{IPR}_{n,\mathbf{k}} = \frac{\int_A \left| u_{n,\mathbf{k}}(\bm{\rho}) \right|^4 d^2\bm{\rho}}{\left[ \int_A \left| u_{n,\mathbf{k}}(\bm{\rho}) \right|^2 d^2\bm{\rho} \right]^2},
\label{IPR}
\end{equation}
where $A = L^2$ is the area of our disordered system. As can be easily deduced from equation\ (\ref{IPR}), a mode extended over the whole system will have IPR $\sim 1/A$, whereas a localized mode will have a larger IPR. We define the localization length $\xi_{n,\mathbf{k}}$ of a mode as
\begin{equation}
\xi_{n,\mathbf{k}} = \frac{1}{2}\sqrt{\frac{1}{\mathrm{IPR}_{n,\mathbf{k}}}},
\label{LL}
\end{equation}
where the numerical prefactor is somewhat arbitrary and will be of no importance in the following because we will compare $\xi_{n,\mathbf{k}}$ for different disorder strengths and system sizes but will not base any important conclusion on the magnitude of $\xi_{n,\mathbf{k}}$.

In a disordered system, $\xi_{n,\mathbf{k}}$ are random quantities and only their statistical properties make physical sense. To obtain the average localization length $\langle \xi(\omega) \rangle$ as a function of frequency, we employ a procedure illustrated in figure\ \ref{fig_method}. A large number of $\xi_{n,\mathbf{k}}$ obtained for 50 different realizations of disorder and corresponding to frequencies $\omega_n(\mathbf{k})$ within a narrow interval around $\omega$, are averaged to obtain $\langle \xi(\omega) \rangle$ which, in its turn, is then fitted by a polynomial function. We will represent our results by the latter in the following.

Whereas the localization length characterizes the spatial extent of eigenmodes, the statistical properties of the spectrum of a disordered system can be characterized by the level spacing statistics $P(s)$, as we already discussed in the Introduction. To compute $P(s)$, we define a dimensionless spacing between adjacent bands $n$ and $n + 1$:
\begin{equation}
s_{n,\mathbf{k}} = \frac{\Delta \omega_{n}(\mathbf{k})}{\langle \Delta \omega_{n}(\mathbf{k}) \rangle},
\label{s_parameter}
\end{equation}
where $\Delta \omega_{n}(\mathbf{k}) = \omega_{n+1}(\mathbf{k}) - \omega_{n}(\mathbf{k})$. Examples of $P(s)$ obtained for TM modes are shown in figure\ \ref{statistics_k}. Previous studies of $P(s)$ in photonic \cite{LNG2002,JJC2013} and electronic \cite{ERM1994} crystals without disorder have shown that the degeneracy of the spectrum in symmetry points $\Gamma$, $X$ or $M$ (see figure \ref{fig_band_structure}) can have a strong influence on $P(s)$ by suppressing the phenomenon of level repulsion. This suppression can be erroneously interpreted as a signature of localized modes. We observe remnants of this phenomenon at weak disorder $\alpha \leq 0.2$ (see figure\ \ref{statistics_k}a). To avoid the confusion between the phenomena due to the high symmetry of the perfect crystal from which we derive our disordered system and the localized nature of eigenmodes, for weak disorder $\alpha \leq 0.2$ we will use only $s_{n,\mathbf{k}}$ for $\mathbf{k}$ near the center of the irreducible Brillouin zone. More precisely, for $\alpha \leq 0.2$, $P(s)$ will be calculated using $s_{n,\mathbf{k}}$ with $k$ within an interval of width $\pi/2L$ around $k = \pi/2L$. In contrast, for $\alpha > 0.2$, we will use $s_{n,\mathbf{k}}$ corresponding to all $k \in [0, \pi/L]$ because in this case $P(s)$ turns out to be independent of $k$ within statistical errors of our analysis, as illustrated by figures\ \ref{statistics_k}b and \ref{statistics_k}c (similar results were obtained for TE modes, but we do not present them here). Finally, very low frequencies $\omega_n(k)a/2\pi c < 0.1$ will be eliminated from the analysis to reduce finite size effects. Such low frequencies correspond to long wavelengths $\lambda > 10a$ (wavelength in the air), which are comparable with the size of our disordered system $L$. For TE modes we restrict the analysis to even larger $\omega_n(k)a/2\pi c > 0.35$ which is made possible by a higher position of the most interesting frequency range---the one corresponding to a partial spectral gap shown in pink in figure \ref{fig_band_structure}b---than for TM modes.

\section*{\label{sec:results}Results}

\subsection*{\label{sec:loc_length}Localization length}

\begin{figure*}[t]
\centering
\includegraphics[width=0.9\textwidth]{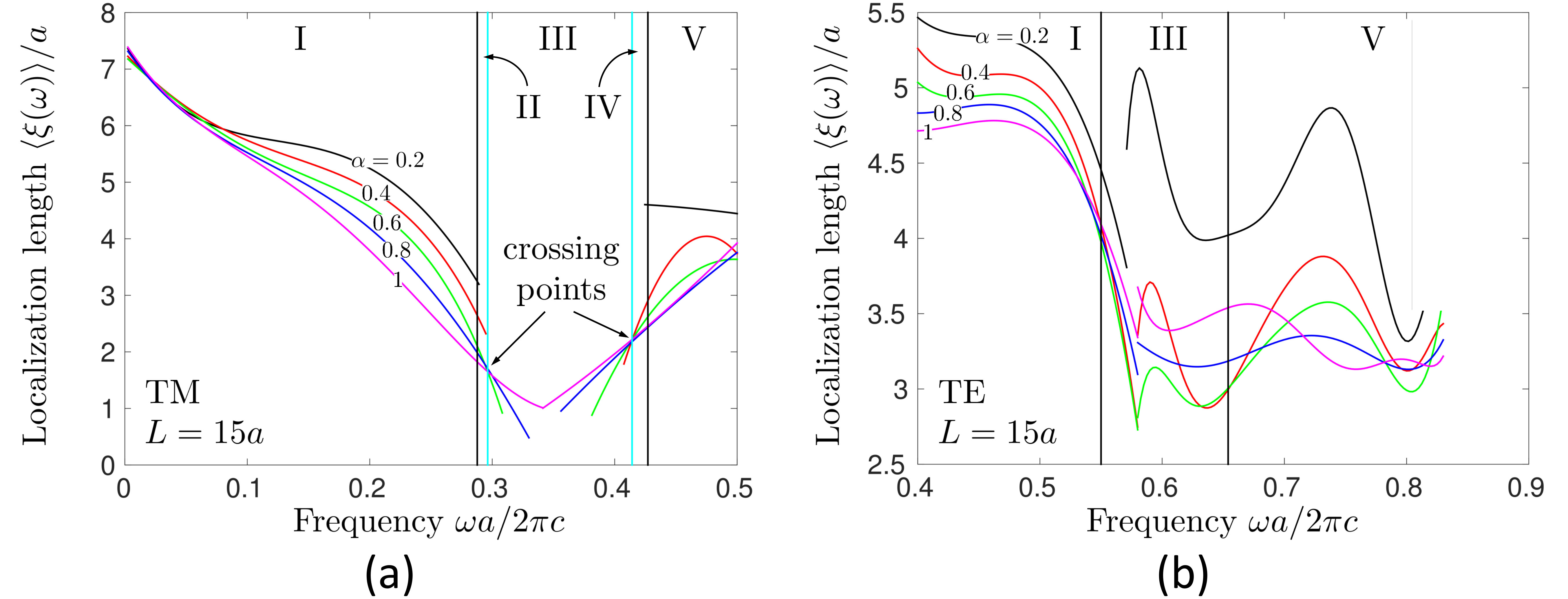}
\caption{Average localization lengths for TM (a) and TE (b) modes for different values of randomness $\alpha = 0.2$--1. In the panel (a), black vertical lines show band edges in the absence of randomness (i.e. for $\alpha = 0$) whereas cyan vertical lines show the positions of points where lines corresponding to different $\alpha$ cross. In the panel (b), black vertical lines show the edges of a partial spectral gap existing for the propagation in the $\Gamma$X direction. The vertical lines shown in this figure divide the spectrum of our system in frequency bands (5 for TM modes and 3 for TE modes, marked by Greek numbers I--V) in which the behavior of the system is found to be different.}
\label{fig_xi}
\end{figure*}

\begin{figure*}[h!]
\centering
\includegraphics[width=0.9\textwidth]{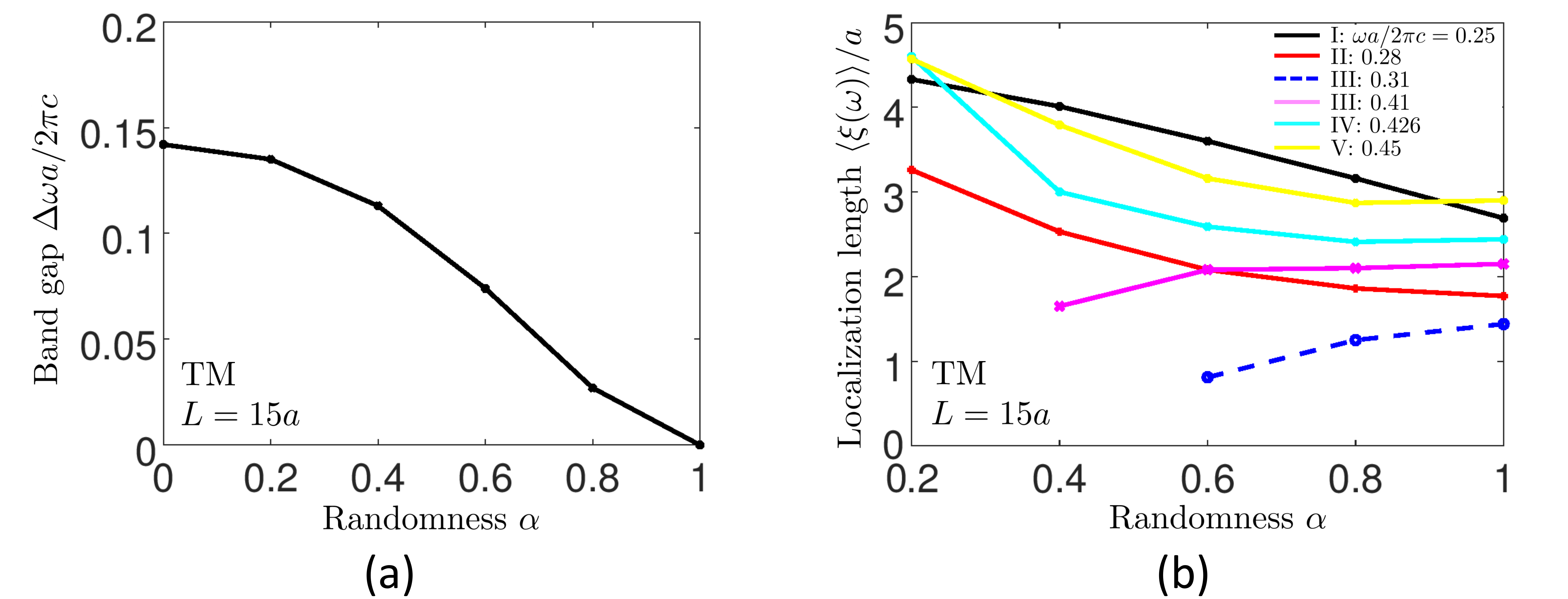}
\caption{Closing of the band gap for TM modes with increasing randomness $\alpha$ (a) and the typical dependencies of the localization lengths of TM modes at frequencies in ranges I--V defined in figure \ref{fig_xi}a (b).}
\label{fig_bg_ll}
\end{figure*}

Figure \ref{fig_xi} shows the frequency dependencies of the average localization lengths $\langle \xi(\omega) \rangle$ for TM (figure\ \ref{fig_xi}a) and TE (figure\ \ref{fig_xi}b) modes, calculated for different disorder strengths $\alpha$. Several important differences are obvious between TM and TE cases. First, $\langle \xi(\omega) \rangle$ reaches smaller values for TM than for TE modes, with the shortest localization lengths achieved inside the bandgap of the initial periodic structure. Second, the localization length of TM modes decays towards the center of the bandgap from both sides, whereas for TE modes, such a decay is clearly visible only from the low-frequency side, the localization length exhibiting a complicated oscillatory behavior on the high-frequency side of the partial bandgap. An interesting feature that we observe in figure\ \ref{fig_xi}a is the crossing of lines corresponding to different $\alpha$ at frequencies that are close to band edges but do not coincide with them, lying slightly inside the bandgap of the structure without disorder. We marked the approximate positions of these crossings by cyan vertical lines in figure\ \ref{fig_xi}a. Together with the band edges of the ideal crystal without disorder, the latter split the spectrum of TM modes in 5 different regions: the frequencies below (I) and above (V) the band gap of the ideal crystal, the middle of the bandgap (III), and the frequencies near band edges (II and IV). Analysis of TE modes (figure\ \ref{fig_xi}b) does not suggest the same fine structure in this case, leaving us with 3 different regions that can be clearly identified: the frequencies below (I) and above (V) the partial bandgap, and the partial bandgap itself (III). We summarize this classification of spectral regions in table \ref{tab1}. Note that the emergence of electromagnetic modes in the spectral regions II--IV leads to progressive narrowing of the gap in the spectrum of TM modes with increasing $\alpha$ and ultimately to its complete closing for $\alpha = 1$, as we illustrate in figure \ref{fig_bg_ll}a. This figure also illustrates the role of periodicity in the formation of the first band gap in the photonic crystal that we consider. It is well known that coupling between scattering (Mie) resonances of dielectric cylinders can give rise to gaps in the electromagnetic spectrum of a structure that they form, even when the structure is not periodic \cite{lido98,rock06,florescu09,froufe17}. The TM band gap that we consider is certainly also sensitive to the properties (dielectric constant, radius) of individual cylinders, but closes when sufficient disorder is introduced and thus would not exist in a fully random structure.

\begin{table}
\centering
\caption{\label{tab1}Ranges of frequencies $\omega a/2\pi c$ defined in figure\ \ref{fig_xi}.}
\begin{tabular}{lll}
\hline
Polarization & TM & TE \\
\hline
I   & 0.1--0.28  & 0.35--0.55 \\
II  & 0.28--0.29 &  \\
III & 0.29--0.41 & 0.55--0.66\\
IV  & 0.41--0.43 & \\
V   & 0.43--0.5  & 0.66--0.83\\
\hline
\end{tabular}
\end{table}

For frequencies outside the range delimited by cyan lines in figure \ref{fig_xi}a (spectral regions I, II, IV and V), $\langle \xi(\omega) \rangle$ exhibits a behavior that could be expected because stronger disorder (larger $\alpha$) yields shorter $\langle \xi(\omega) \rangle$, see figure \ref{fig_bg_ll}b. However, inside the frequency band delimited by cyan lines (spectral region III), the dependence of $\langle \xi(\omega) \rangle$ on $\alpha$ is just the opposite: larger $\alpha$ (i.e. stronger disorder) gives longer $\langle \xi(\omega) \rangle$ (blue dashed and magenta solid lines in figure \ref{fig_bg_ll}b). We believe that such a behavior is due to the different localization mechanisms at work for modes with frequencies in the middle of the gap and near band edges. On the one hand, the modes with frequencies in the middle of the gap (region III in figure\ \ref{fig_xi}a) need disorder to appear (because there are strictly no modes in the gap in the absence of disorder), but their localized nature relies on the periodic structure of the system. In other words, they benefit from Bragg reflections from the parts of the photonic crystal surrounding them. Increasing disorder makes these modes more numerous and hence easier to detect, but reduces their spatial localization making their localization lengths larger because of less efficient Bragg reflections. On the other hand, the modes near band edges (regions II and IV in figure\ \ref{fig_xi}a) get more and more localized when the randomness $\alpha$ is increased. Stronger disorder (i.e., larger $\alpha$) provides a shorter localization length for this modes, as expected for the genuine Anderson localization. These two different ``modes of localization'' have been previously predicted to exist in 1D random periodic-on-average systems \cite{deych98}.

\begin{figure*}
\centering
\includegraphics[width=0.8\textwidth]{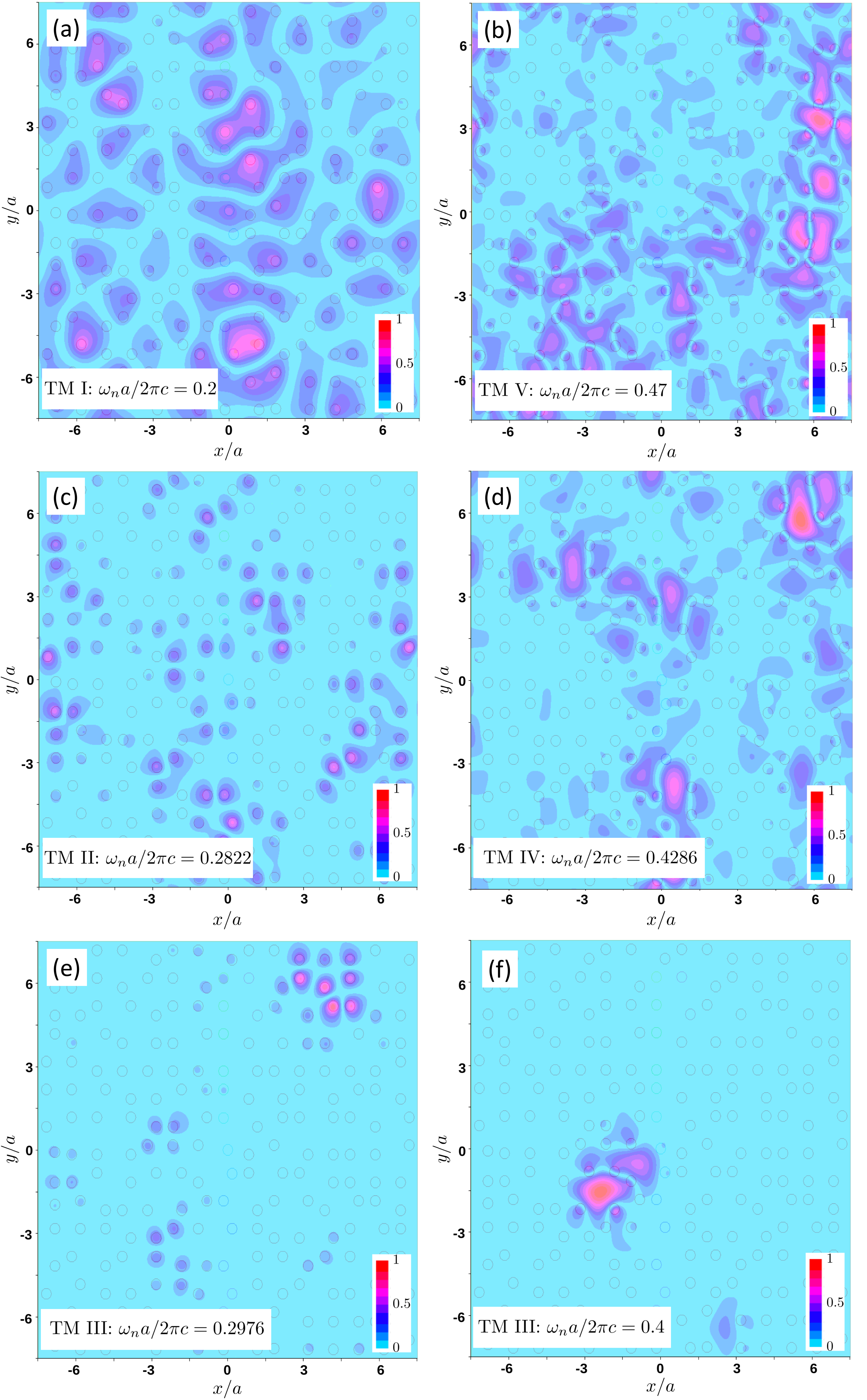}
\caption{Typical intensity patterns $|u_{n,\mathbf{k}}(\bm{\rho})|^2$ of TM modes for $k = \pi/L$ and $\alpha = 0.6$ in different spectral regions: I --- below the band gap (a), V --- above the bandgap (b), II and IV --- near band edges (c, d), III --- inside the bandgap (e,f). $|u_{n,\mathbf{k}}(\bm{\rho})|^2$ was normalized to have a maximum value of 1 in each panel.}
\label{fig_mode}
\end{figure*}

\begin{figure*}
\centering
\includegraphics[width=0.9\textwidth]{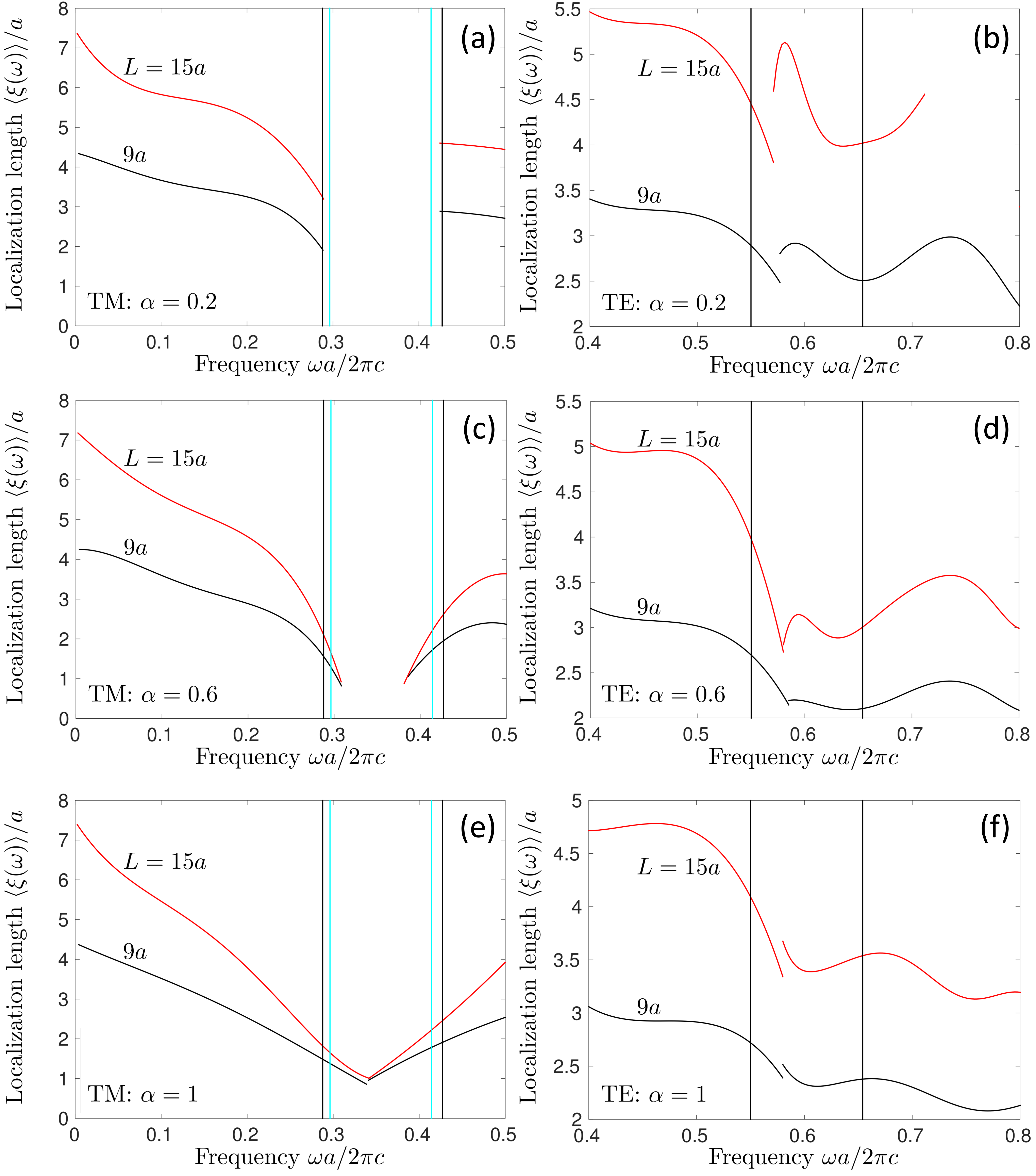}
\caption{Average localization length $\langle \xi(\omega) \rangle$ for TM [first row, panels (a), (c), and (e)] and TE [second row, panels (b), (d), and (f)] modes at different randomness $\alpha = 0.2$ (a, b), 0.6 (c, d), and 1 (e, f) and for two different system sizes $L = 9a$ (black lines) and 15a (red lines). Vertical lines are the same as in figures\ \ref{fig_xi}a and \ref{fig_xi}b.}
\label{fig_xi_tm_te}
\end{figure*}

Figure \ref{fig_mode} illustrates the intensity patterns of typical TM modes in different frequency ranges. No difference can be identified by eye between the modes with frequencies below (figure\ \ref{fig_mode}a) or above (figure\ \ref{fig_mode}b) the band gap, which both look extended. The modes with frequencies inside the band gap of the periodic structure but near band edges (frequency ranges II and IV, see figures\ \ref{fig_mode}c and \ref{fig_mode}d) also look extended and only the modes with frequencies inside the range III appear as clearly localized in space (figures\ \ref{fig_mode}e and \ref{fig_mode}f). This may seem to be in a contradiction with figure\ \ref{fig_xi}a from which we see that the modes shown in figures\ \ref{fig_mode}c and \ref{fig_mode}e, as well as those in figures\ \ref{fig_mode}d and \ref{fig_mode}f, have approximately the same localization length $\xi \approx 2a$ on average. This apparent contradiction illustrates the ambiguity of our definition (\ref{LL}) of the localization length $\xi$ via the IPR. Indeed, the IPR and $\xi$ characterize, respectively, the inverse area and the spatial extent of a region within which the mode intensity is considerable but they do not change if we split the area in $M$ parts of size $\xi/\sqrt{M}$ that we separate in space by arbitrary large distances. Once we realize this, it becomes easier to reconcile the different appearance of the modes shown in figures\ \ref{fig_mode}c and \ref{fig_mode}e (or in figures\ \ref{fig_mode}d and \ref{fig_mode}f) with very similar values of IPR and $\xi$ that they have. Thus, the localization length $\xi$ defined by equation\ (\ref{LL}) cannot be taken as the only measure of localization and other quantities, such as, e.g., the level spacing statistics that we will analyze below, have to be considered.

Figures \ref{fig_mode}c and  \ref{fig_mode}d and, especially, \ref{fig_mode}e and \ref{fig_mode}f illustrate an important difference in the spatial structure of disorder-induced modes. The modes with frequencies near the lower edge of the band gap tend to be localized inside the dielectric cylinders whereas the modes with frequencies near the upper band edge---in the space (air) between the cylinders. Disorder-induced modes inherit this behavior from the modes of the photonic crystal without disorder, where the corresponding ``dielectric'' (for frequencies below the band gap) and ``air'' (for frequencies above the band gap) bands are known to exist \cite{joan08}.

Numerical studies of localization phenomena, including the one reported here, are subject to limitations due to the finite size $L$ of disordered regions that can be analyzed ($L \leq 15a$ in our case). It is often difficult to decide whether a mode with a localization length $\xi$ equal to a fraction of $L$ should be considered spatially localized or not. To resolve this difficulty, one usually studies the evolution of $\xi$ with $L$. $\xi$ (roughly) proportional to $L$ indicates that, most probably, the mode is extended and the surface that it will effectively cover in the limit of $L \to \infty$ will be of order $L^2$. Localized modes distinguish themselves by a localization length $\xi$ that is (almost) independent of $L$, suggesting that the mode will cover a negligible part of the surface of the disordered region in the limit of $L \to \infty$. A practical realization of this analysis is illustrated in figure\ \ref{fig_xi_tm_te}. This figure witnesses that localized modes appear for sufficiently strong disorder $\alpha \gtrsim 0.4$ in the central part of the band gap where the localization lengths computed for two different sizes of the disordered region $L = 9a$ and $L = 15a$, roughly coincide for TM polarization. When disorder is weak (e.g., $\alpha = 0.2$ in figure\ \ref{fig_xi_tm_te}a) or for frequencies outside the band gap at any $\alpha$, the average localization length $\langle \xi(\omega) \rangle$ behaves as if the modes were extended. The same is true for TE polarization at any $\alpha$ and throughout the considered frequency range. Because for TE modes, $\langle \xi(\omega) \rangle$ increases with $L$ in a roughly uniform way throughout the spectrum, we cannot conclude about the eventual localization of these modes in the limit $L \to \infty$. It may happen that the modes eventually get localized when $L$ is increased and that their apparent extended nature under conditions of figures\ \ref{fig_xi_tm_te}b, \ref{fig_xi_tm_te}d and \ref{fig_xi_tm_te}f is due to the fact that their localization lengths are larger than $L$. In any case, however, the analysis of figure\ \ref{fig_xi_tm_te} indicates that TM modes are easier to localize by disorder than TE ones, which under certain experimental conditions may lead to TE modes that can be considered extended for at least some practical purposes. This conclusion, however, is restricted to a structure composed of high-index dielectric cylinders in a low-index material (air). In an inverted structure (cylindrical holes in a high-index bulk material), TE modes may be stronger localized than TM ones \cite{faggiani16,garcia12,garcia13}.

\subsection*{\label{sec:lev_spac_statistics}Level-spacing statistics}

\begin{figure*}[t]
\centering
\includegraphics[width=0.9\textwidth]{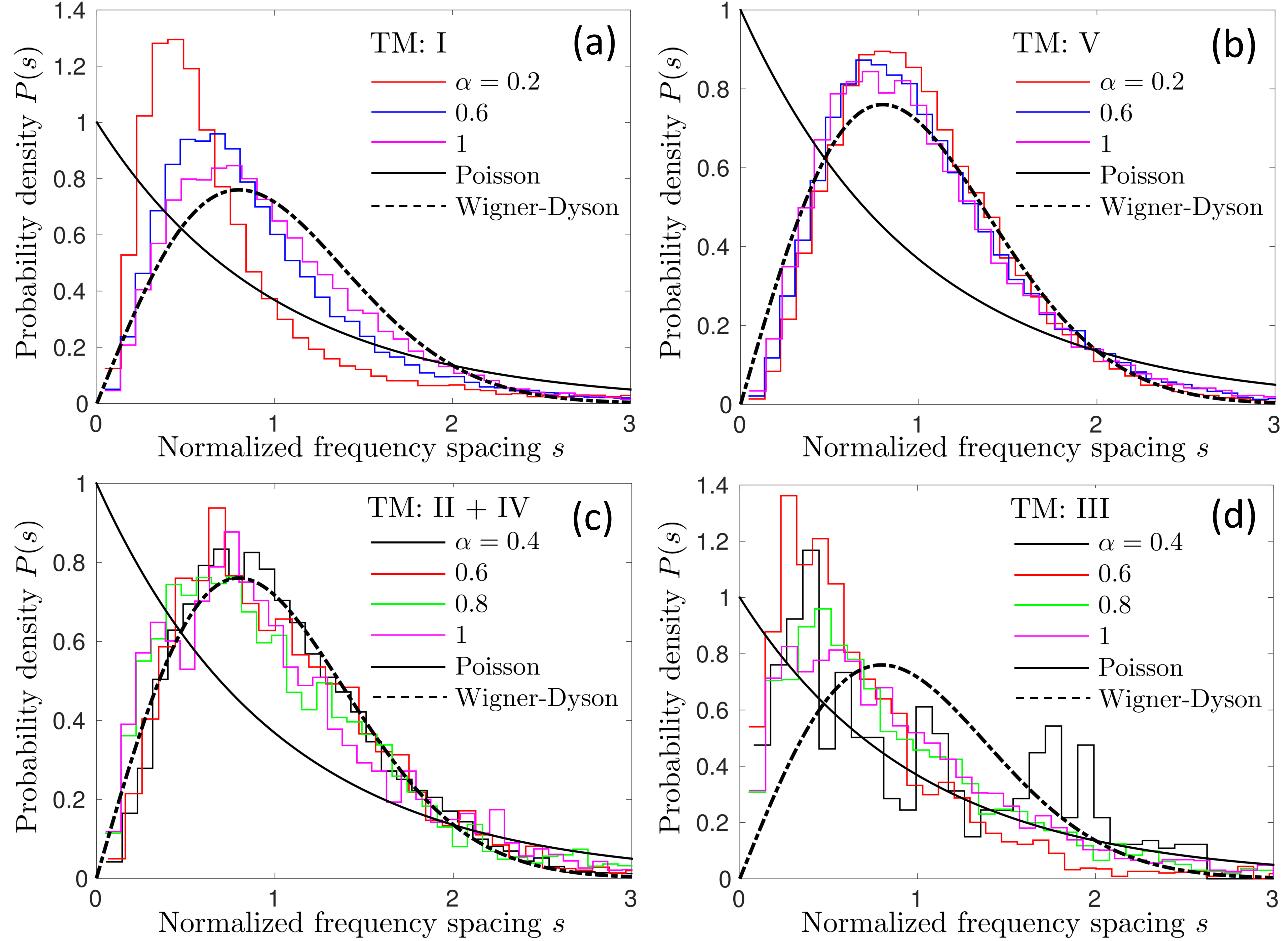}
\caption{Probability distributions of normalized eigenfrequency spacings for TM modes in different parts of the spectrum: I --- below bandgap (a), V --- above bandgap (b), II + IV --- inside bandgap near band edges (c) and III --- in the middle of the bandgap (d).}
\label{fig_spacing_tm}
\end{figure*}

\begin{figure*}[t]
\centering
\includegraphics[width=0.9\textwidth]{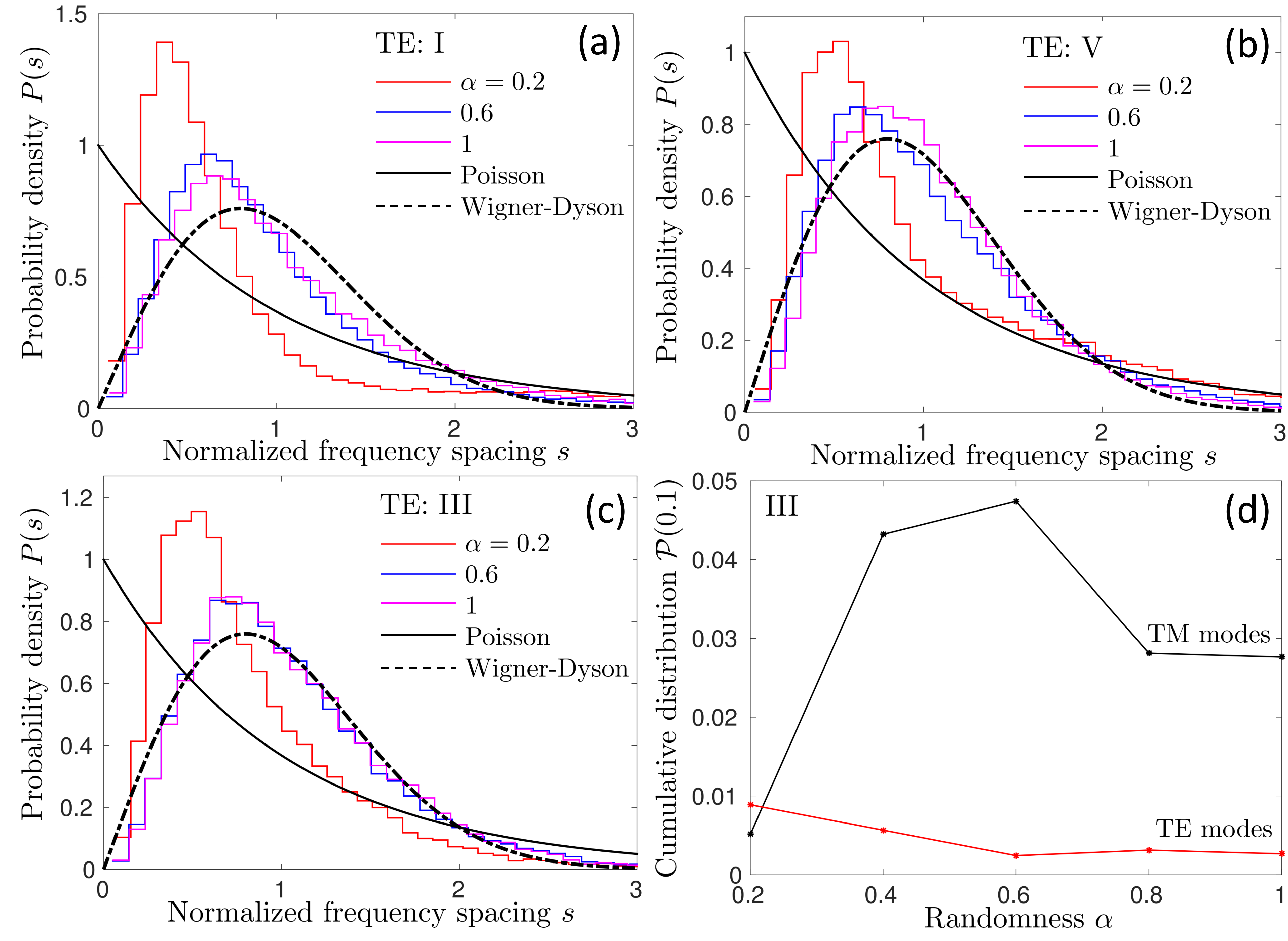}
\caption{Probability distributions of normalized eigenfrequency spacings for TE modes in different parts of the spectrum: I --- below the pseudo gap (a), V --- above the pseudo gap (b), III --- inside the pseudo gap (c). Panel (d) shows the probability of having $s < 0.1$, which can serve as a measure of eigenfrequency repulsion, for TM and TE modes as a function of randomness $\alpha$. }
\label{fig_spacing_te}
\end{figure*}

Now that we have characterized the localization properties of the modes of our disordered photonic crystal, we are ready to study its level spacing statistics $P(s)$ in a meaningful way. Figure \ref{fig_spacing_tm} shows $P(s)$ computed for TM modes in frequency ranges defined in table \ref{tab1}. For frequencies $\omega_n(k)$ below the band gap, $P(s)$ tends to the Wigner-Dyson distribution (\ref{P_WD}) when the disorder is increased, see figure\ \ref{fig_spacing_tm}a. The same is true for $\omega_n(k)$ above the band gap where a good agreement of numerical results with equation\ (\ref{P_WD}) is found for all $\alpha \geq 0.2$ (figure\ \ref{fig_spacing_tm}b). This reflects the chaotic nature of our disordered photonic crystal in the classical limit and signals that the modes are extended both below and above the band gap, exhibiting a strong phenomenon of level repulsion: $P(0) = 0$. A very similar behavior is found for disorder-induced modes inside the band gap but close to its edges (figure\ \ref{fig_spacing_tm}c). In contrast, $P(s)$ computed for the modes in the middle of the gap, is closer to the Poisson distribution (\ref{P_P}), with the phenomenon of level repulsion being suppressed: $P(0) > 0$, see figure\ \ref{fig_spacing_tm}d. This is in agreement with our previous conclusion about the localized character of modes in the spectral region III.

The level spacing distribution $P(s)$ for TE modes is shown in figure\ \ref{fig_spacing_te}. $P(s)$ is close to the Wigner-Dyson result (\ref{P_WD}) in all frequency ranges, including the partial band gap region III (figure \ref{fig_spacing_te}c), provided that the disorder is strong enough ($\alpha \gtrsim 0.6$). This is in agreement with the extended nature of eigenmodes following from the analysis of their localization lengths. For weak disorder $\alpha \lesssim 0.2$, $P(s)$ differs from the Wigner-Dyson distribution and has its maximum shifted towards smaller $s$. We attribute this to the remnants of the symmetries of the crystal without disorder which are still partially present at small $\alpha$. Even if the shape of $P(s)$ for TE modes at small $\alpha$ differs from the Wigner-Dyson distribution, $P(s)$ exhibits the phenomenon of level repulsion at all $\alpha$. To illustrate this fact, figure\ \ref{fig_spacing_te}d shows the cumulative distribution function of $s$ defined as
\begin{equation}
{\cal P}(s) = \int\limits_0^{s} P(s') ds',
\label{p01}
\end{equation}
for a small value of $s = 0.1$. ${\cal P}(s)$ is equal to the probability of having an eigenfrequency spacing inferior to $s$ and can serve as a measure of level repulsion if we take $s$ sufficiently small. We see from figure\ \ref{fig_spacing_te}d that for TE modes, the level repulsion remains weak and roughly the same ${\cal P}(0.1) \simeq 0.05$ independent of the strength of disorder $\alpha$. In contrast, the disorder considerably weakens the level repulsion for TM modes and increases ${\cal P}(0.1)$ by an order of magnitude when $\alpha$ grows from 0.2 to 0.6. Once again, this observation is consistent with our previous conclusions based on the analysis of the localization lengths of TM and TE modes. Note that the suppression of level repulsion weakens for strong disorder $\alpha > 0.6$. We relate this phenomenon with the reduction of Bragg scattering and the weakening of the associated Bragg-scattering-assisted localization mechanism at strong disorder.

\section*{\label{sec:conclusion}Conclusions}

We performed a detailed statistical analysis of eigenmodes of a 2D disordered photonic crystal composed of parallel, circular dielectric (silicon) rods in air. Dependencies of the average localization lengths of eigenmodes on their frequencies and on the size of the disordered system were analyzed. In the TM case (electric field parallel to the rods), the modes arising in the middle of the band gap of the ideal crystal are spatially localized due to Bragg reflections on the almost periodic structure surrounding them. The localization length of these modes grows with increasing disorder. In contrast, the modes appearing near band edges have a tendency to become more and more localized when the disorder is increased (genuine Anderson localization), although we cannot claim with certainty that these modes are spatially localized in our relatively small samples (up to $15 \times 15$ rods). In the TE case (magnetic field parallel to the rods), the band gap of the ideal photonic crystal in the considered frequency range is only partial and no clear sign of localized modes were identified. This indicates that TM modes are easier to get localized by disorder than TE ones, although the conclusion may be different in an inverted structure (cylindrical air holes in a high-index dielectric material).

For TM modes outside the band gap and TE modes at all frequencies, the level spacing statistics is found to be close to the Wigner-Dyson distribution, typical for classically chaotic systems and for disordered wave systems with extended eigenmodes. Significant deviations from the Wigner-Dyson distribution are observed at weak disorder and are attributed to the residual symmetries due to the periodicity of the initial photonic crystal. The latter symmetries partially suppress the level repulsion phenomenon and modify the overall shape of the distribution. For TM modes inside the band gap of the ideal crystal, the phenomenon of level repulsion is significantly suppressed and the level spacing statistics approaches the Poisson distribution, as expected for statistically independent eigenfrequencies corresponding to spatially localized eigenmodes. The level repulsion weakens upon increasing disorder, reaches a minimum at some optimal disorder for which the interplay between the Bragg-scattering-assisted and disorder-induced localization is optimal, and then starts to strengthen again due to the suppression of Bragg scattering by disorder.

An interesting extension of this work may concern a comparison of results obtained here for a system of dielectric rods in air with those for an inverted structure in which circular holes are drilled in a dielectric material. The latter structure provides stronger scattering for TE modes and such a comparison can therefore give an insight into the role of polarization in the phenomenon of Anderson localization of light.

\section*{Acknowledgements}

This work was funded by the Agence Nationale de la Recherche (project ANR-14-CE26-0032 LOVE). All the computations presented in this paper were performed using the Froggy platform of the CIMENT infrastructure (https://ciment.ujf-grenoble.fr), which is supported by the Rhone-Alpes region (grant CPER07\verb!_!13 CIRA) and the Equip@Meso project (reference ANR-10-EQPX-29-01) of the programme Investissements d'Avenir supervised by the Agence Nationale de la Recherche.

\section*{Author contributions statement}

S.E.S. formulated the problem. J.M.E. performed the calculations. Both authors contributed to the discussion of results and writing the paper.

\section*{Additional information}

\textbf{Competing interests.}
The authors declare no competing interests.


\begin{thebibliography}{99}

\bibitem{wigner51}
Wigner, E.P.
On the Statistical Distribution of the Widths and Spacings of Nuclear Resonance Levels.
\textit{Proc. Cambridge Philos. Soc.} \textbf{47}, 790--798 (1951).

\bibitem{bohigas83}
Bohigas, O., Haq, R.U. \& Pandey, A.
Fluctuation Properties of Nuclear Energy Levels and Widths: Comparison of Theory with Experiment.
In: Nuclear Data for Science and Technology. Edited by B\"{o}ckhoff, K.H. (Reidel, Dordrecht, Netherlands, 1983), pp. 809--813.

\bibitem{mehta67}
Mehta, M.L. Random Matrices and the Statistical Theory of Energy Levels
(Academic, New York, 1967).

\bibitem{izrailev90}
Izrailev, F.M.
Simple Models of Quantum Chaos: Spectrum and Eigenfunctions.
\textit{Phys. Rep.} \textbf{196}, 299--392 (1990).

\bibitem{guhr98}
Guhr, T.,  Mu\"{u}ller-Groeling, A. \& Weidenm\"{u}ller, H.A.
Random-Matrix Theories in Quantum Physics: Common Concepts.
\textit{Phys. Rep.} \textbf{299}, 189--425 (1998).

\bibitem{haake01}
Haake, F.
Quantum Signatures of Chaos. 2nd ed. (Springer, Berlin, 2001).

\bibitem{anderson58}
Anderson, P.W.
Absence of Diffusion in Certain Random Lattices.
\textit{Phys. Rev.} \textbf{109}, 1492--1505 (1958).

\bibitem{lagendijk09}
Lagendijk, A., Van Tiggelen, B.A. \& Wiersma, D.S.
Fifty Years of Anderson Localization.
\textit{Phys. Today} \textbf{62}(8), 24--29 (2009).

\bibitem{kramer93}
Kramer, B. \& MacKinnon, A.
Localization: Theory and Experiment.
\textit{Rep. Prog. Phys.} \textbf{56}, 1469--1564 (1993).

\bibitem{billy08}
Billy, J., Josse, V., Zuo, Z., Bernard, A., Hambrecht, B., Lugan, P., Cl\'{e}ment, D., Sanchez-Palencia, L., Bouyer, P. \& Aspect, A.
Direct Observation of Anderson Localization of Matter Waves in a Controlled Disorder.
\textit{Nature} \textbf{453}, 891--894 (2008).

\bibitem{chabe08}
Chab\'{e}, J., Lemari\'{e}, G., Gr\'{e}maud, B., Delande, D., Szriftgiser, P. \& Garreau, J.C.
Experimental Observation of the Anderson Metal-Insulator Transition with Atomic Matter Waves.
\textit{Phys. Rev. Lett.} \textbf{101}, 255702 (2008).

\bibitem{jendr12}
Jendrzejewski, F., Bernard, A., M\"{u}ller, K., Cheinet, P., Josse, V., Piraud, M., Pezz\'{e}, L.,	Sanchez-Palencia, L., Aspect, A. \& Bouyer, P.
Three-Dimensional Localization of Ultracold Atoms in an Optical Disordered Potential.
\textit{Nature Phys.} \textbf{8}, 398--403 (2012).

\bibitem{hu08}
Hu, H., Strybulevych, A., Page, J.H., Skipetrov, S.E. \& Van Tiggelen, B.A.
Localization of Ultrasound in a Three-Dimensional Elastic Network.
\textit{Nature Phys.} \textbf{4}, 945--948 (2008).

\bibitem{segev13}
Segev, M., Silberberg, Y. \& Christodoulides, D.N.
Anderson Localization of Light.
\textit{Nature Photon.} \textbf{7}, 197--204 (2013).

\bibitem{wiersma13}
Wiersma, D.S.
Disordered Photonics.
\textit{Nat. Photon.} \textbf{7}, 188--196 (2013).

\bibitem{redding13}
Redding, B., Liew, S.F., Sarma, R. \& Cao, H.
Compact Spectrometer Based on a Disordered Photonic Chip.
\textit{Nat. Photon.} \textbf{7}, 746--751 (2013).

\bibitem{hsieh15}
Hsieh, P., Chung, C., McMillan, J.F., Tsai, M., Lu, M., Panoiu, N.C. \& Wong C.W.
Photon Transport Enhanced by Transverse Anderson Localization in Disordered Superlattices.
\textit{Nat. Phys.} \textbf{11}, 268--274 (2015).

\bibitem{trojak17}
Trojak, O.J., Crane, T. \& Sapienza, L.
Optical Sensing with Anderson-Localised Light.
\textit{Appl. Phys. Lett.} \textbf{111}, 141103 (2017).

\bibitem{john84}
John, S.
Electromagnetic Absorption in a Disordered Medium near a Photon Mobility Edge.
\textit{Phys. Rev. Lett.} \textbf{53}, 2169--2172 (1984).

\bibitem{anderson85}
Anderson, P.W.
The Question of Classical Localization: A Theory of White Paint?
\textit{Philos. Mag. B} \textbf{52}, 505--509 (1985).

\bibitem{john87}
John, S.
Strong Localization of Photons in Certain Disordered Dielectric Superlattices.
\textit{Phys. Rev. Lett.} \textbf{58}, 2486--2489 (1987).

\bibitem{john91}
John, S.
Localization of Light.
\textit{Phys. Today} \textbf{44}(5), 32--40 (1991).

\bibitem{vantiggelen94}
Van Tiggelen, B.A. \& Kogan, E.
Analogies Between Light and Electrons: Density of States and Friedel's Identity.
\textit{Phys. Rev. A} \textbf{49}, 708--713 (1994).

\bibitem{berry97}
Berry, M.V. \& Klein, S.
Transparent Mirrors: Rays, Waves and Localization.
\textit{Eur. J. Phys.} \textbf{18}, 222--228 (1997).

\bibitem{chabanov00}
Chabanov, A.A., Stoytchev, M. \& Genack, A.Z.
Statistical signatures of photon localization.
\textit{Nature} \textbf{404}, 850--853 (2000).

\bibitem{schwartz07}
Schwartz, T., Bartal, G., Fishman, S. \& Segev, M.
Transport and Anderson Localization in Disordered Two-Dimensional Photonic Lattices.
\textit{Nature} \textbf{446}, 52--55 (2007).

\bibitem{wiersma97}
Wiersma, D.S., Bartolini, P., Lagendijk, A. \& Righini, R.
Localization of Light in a Disordered Medium.
\textit{Nature} \textbf{390}, 671--673 (1997).

\bibitem{vanderbeek12}
Van der Beek, T., Barthelemy, P., Johnson, P.M., Wiersma, D.S. \& Lagendijk, A.
Light Transport Through Disordered Layers of Dense Gallium Arsenide Submicron Particles.
\textit{Phys. Rev. B} \textbf{85}, 115401 (2012).

\bibitem{sperling13}
Sperling, T., B\"{u}hrer, W., Aegerter, C.M. \& Maret, G.
Direct Determination of the Transition to Localization of Light in Three Dimensions.
\textit{Nature Photon.} \textbf{7}, 48--52 (2013).

\bibitem{sperling16}
Sperling, T., Schertel, L., Ackermann, M., Aubry, G.J., Aegerter, C.M. \& Maret, G.
Can 3D light localization be reached in `white paint'?
\textit{New J. Phys.} \textbf{18}, 013039 (2016).

\bibitem{skip16}
Skipetrov, S.E. \& Page, J.H.
Red Light for Anderson Localization.
\textit{New J. Phys.} \textbf{18}, 021001 (2016).

\bibitem{faez11}
Faez, S., Lagendijk, A. \& Ossipov, A.
Critical Scaling of Polarization Waves on a Heterogeneous Chain of Resonators.
\textit{Phys. Rev. B} \textbf{83}, 075121 (2011).

\bibitem{maximo15}
Maximo, C.E., Piovella, N., Courteille, Ph.W., Kaiser, R. \& Bachelard, R.
Spatial and Temporal Localization of Light in Two Dimensions.
\textit{Phys. Rev. A} \textbf{92}, 062702 (2015).

\bibitem{skip14}
Skipetrov, S.E. \& Sokolov, I.M.
Absence of Anderson Localization of Light in a Random Ensemble of Point Scatterers.
\textit{Phys. Rev. Lett.} \textbf{112}, 023905 (2014).

\bibitem{bellando14}
Bellando, L., Gero, A., Akkermans, E. \& Kaiser, R.
Cooperative Effects and Disorder: A Scaling Analysis of the Spectrum of the Effective Atomic Hamiltonian.
\textit{Phys. Rev. A} \textbf{90}, 063822 (2014).

\bibitem{escalante17}
Escalante, J.M. \& Skipetrov, S.E.
Longitudinal Optical Fields in Light Scattering from Dielectric Spheres and Anderson Localization of Light.
\textit{Ann. Phys. (Berlin)} \textbf{529}, 1700039 (2017).

\bibitem{sapienza10}
Sapienza, L., Thyrrestrup, H., Stobbe, S., Garcia, P.D., Smolka, S. \& Lodahl, P.
Cavity Quantum Electrodynamics with Anderson-Localized Modes.
\textit{Science} \textbf{327}, 1352--1355 (2010).

\bibitem{riboli14}
Riboli, F., Caselli, N., Vignolini1, S., Intonti, F., Vynck, K., Barthelemy, P., Gerardino, A., Balet, L., Li, L.H., Fiore, A., Gurioli, M. \& Wiersma, D.S.
Engineering of Light Confinement in Strongly Scattering Disordered Media.
\textit{Nat. Mater.} \textbf{13}, 720--725 (2014).

\bibitem{riboli17}
Riboli, F., Uccheddu, F., Monaco, G., Caselli, N., Intonti, F., Gurioli, M. \& Skipetrov, S.E.
Tailoring Correlations of the Local Density of States in Disordered Photonic Materials.
\textit{Phys. Rev. Lett.} \textbf{119}, 043902 (2017).

\bibitem{crane17}
Crane, T., Trojak, O.J., Vasco, J.P., Hughes, S. \& Sapienza, L.
Anderson Localization of Visible Light on a Nanophotonic Chip.
\textit{ACS Photonics} \textbf{4}, 2274--2280 (2017).

\bibitem{lee18}
Lee, M., Lee, J., Kim, S., Callard, S., Seassal, C. \& Jeon, H.
Anderson Localizations and Photonic Band-Tail States Observed in Compositionally Disordered Platform.
\textit{Sci. Adv.} \textbf{4}, e160279 (2018).

\bibitem{sigalas96}
Sigalas, M.M., Soukoulis, C.M., Chan, C.-T. \& Turner, D.
Localization of Electromagnetic Waves in Two-Dimensional Disordered Systems.
\textit{Phys. Rev. B} \textbf{53}, 8340--8348 (1996).

\bibitem{asatryan99}
Asatryan, A.A., Robinson, P.A., Botten, L.C., McPhedran, R.C., Nicorovici, N.A. \& De Sterke, C.M.
Effects of Disorder on Wave Propagation in Two-Dimensional Photonic Crystals.
\textit{Phys. Rev. E} \textbf{60}, 6118--6127 (1999).

\bibitem{vanneste05}
Vanneste, C. \& Sebbah, P.
Localized Modes in Random Arrays of Cylinders.
\textit{Phys. Rev. E} \textbf{71}, 026612 (2005).

\bibitem{froufe17}
Froufe-Perez, L.S., Engel, M., Saenz, J.J. \& Scheffold, F.
Band Gap Formation and Anderson Localization in Disordered Photonic Materials with Structural Correlations.
\textit{Proc. Nat. Acad. Sci.} \textbf{114}, 9570--9574 (2017).

\bibitem{huis12}
Huisman, S.R., Ctistis, G., Stobbe, S., Mosk, A.P., Herek, J.L., Lagendijk, A., Lodahl, P., Vos, W.L. \& Pinkse, P.W.H.
Measurement of a Band-Edge Tail in the Density of States of a Photonic-Crystal Waveguide.
\textit{Phys. Rev. B} \textbf{86}, 155154 (2012).

\bibitem{faggiani16}
Faggiani, R., Baron, A., Zang, X., Lalouat, L., Schulz, S.A., O'Regan, B., Vynck, K., Cluzel, B., De Fornel, F., Krauss, T.F. \& Lalanne, P.
Lower Bound for the Spatial Extent of Localized Modes in Photonic-Crystal Waveguides with Small Random Imperfections.
\textit{Sci. Rep.} \textbf{6}, 27037; 10.1038/srep27037 (2016).

\bibitem{mazoyer09}
Mazoyer, S., Hugonin, J.P. \& Lalanne, P.
Disorder-Induced Multiple Scattering in Photonic-Crystal Waveguides.
\textit{Phys. Rev. Lett.} \textbf{103}, 063903 (2009).

\bibitem{vasco17}
Vasco, J.P. \& Hughes, S.
Statistics of Anderson-Localized Modes in Disordered Photonic Crystal Slab Waveguides.
\textit{Phys. Rev. B} \textbf{95}, 224202 (2017).

\bibitem{garcia17}
Garc\'{i}a, P.D. \& Lodahl, P.
Physics of Quantum Light Emitters in Disordered Photonic Nanostructures.
\textit{Ann. Phys. (Berlin)} \textbf{529}, 1600351 (2017).


\bibitem{LNG2002}
Gumen, L.N., Arriaga, J. \& Krokhin, A.A.
Manifestation of Quantum Chaos in Spectra of 2D Photonic Crystals.
\textit{Physica E} \textbf{13}, 459--462 (2002).

\bibitem{JJC2013}
Cruz-Bueno, J.J., Mendez-Bermudez, J.A. \& Arriaga, J.
Spectral Properties of a Two Dimensional Photonic Crystal with Quasi-Integrable Geometry.
\textit{J. Phys.: Conf. Ser.} \textbf{475}, 012009 (2013).

\bibitem{ERM1994}
Mucciolo, E.R., Capaz, R.B., Altshuler, B.L. \& Joannopoulos, J.D.
Manifestation of Quantum Chaos in Electronic Band Structures.
\textit{Phys. Rev. B} \textbf{50}, 8245--8251 (1994).

\bibitem{edwards80}
Edwards, D.F. \& Ochoa, E.
Infrared Refractive Index of Silicon.
\textit{Appl. Opt.} \textbf{19}, 4130--4131 (1980).

\bibitem{deg13}
Degirmenci, E. \& Landais, P.
Finite Element Method Analysis of Band Gap and Transmission of Two-dimensional Metallic Photonic Crystals at Terahertz Frequencies.
\textit{Appl. Opt.} \textbf{52}, 7367--7375 (2013).

\bibitem{joan08}
Joannopoulos, J.D., Johnson, S.G., Winn, J.N. \& Meade, R.D.
Photonic Crystals: Molding the Flow of Light. 2nd ed. (Princeton Univ. Press, Princeton, 2008).

\bibitem{lourtioz05}
Lourtioz, J.-M., Benisty, H., Berger, V., Gerard, J.-M., Maystre, D. \& Tchelnokov, A.
Photonic Crystals: Towards Nanoscale Photonic Devices (Springer-Verlag, Berlin, 2005).

\bibitem{FEM}
Hecht, F.
New development in FreeFem++.
\textit{J. Numer. Math.} \textbf{20}, 251--265 (2012).

\bibitem{JIA2002}
Jin, J.-M.
The Finite Element Method in Electromagnetics. 3rd ed. (John Wiley \& Sons, Hoboken, NJ, 2014).

\bibitem{lido98}
Lidorikis, E., Sigalas, M.M., Economou, E.N., \& Soukoulis, C.M.
Tight-Binding Parametrization for Photonic Band Gap Materials.
\textit{Phys. Rev. Lett.} \textbf{81}, 1405--1408 (1998).

\bibitem{rock06}
Rockstuhl, C., Peschel, U. \& Lederer, F.
Correlation Between Single-Cylinder Properties and Bandgap Formation in Photonic Structures.
\textit{Opt. Lett.} \textbf{31}, 1741--1743 (2006).

\bibitem{florescu09}
Florescu M., Torquato S. \& Steinhardt P.J.
Designer Disordered Materials with Large, Complete Photonic Band Gaps.
\textit{Proc. Natl. Acad. Sci. USA} \textbf{106}, 20658--20663 (2009).

\bibitem{deych98}
Deych, L.I., Zaslavsky, D. \& Lisyansky, A.A.
Statistics of the Lyapunov Exponent in 1D Random Periodic-on-Average Systems.
\textit{Phys. Rev. Lett.} \textbf{81}, 5390--5393 (1998).

\bibitem{garcia12}
Garc\'{i}a, P.D., Stobbe, S., S\"{o}llner, I. \& Lodahl, P.
Nonuniversal Intensity Correlations in a Two-Dimensional Anderson-Localizing Random Medium.
\textit{Phys. Rev. Lett.} \textbf{109}, 253902 (2012).

\bibitem{garcia13}
Garc\'{i}a, P.D., Javadi, A., Thyrrestrup, H. \& Lodahl, P.
Quantifying the Intrinsic Amount of Fabrication Disorder in Photonic-Crystal Waveguides from Optical Far-Field Intensity Measurements.
\textit{Appl. Phys. Lett.} \textbf{102}, 031101 (2013).

\end{thebibliography}
\end{document}